\documentclass[a4paper,11pt]{article}
\usepackage{ascmac}
\usepackage{graphicx}
\usepackage[title]{appendix}

\newcommand{\siki}[1]{Eq.(\ref{eq:#1})}
\newcommand{\zu}[1]{Fig.\ref{fig:#1}}

\newcommand{\rref}[1]{${}^{\mathrm{#1)}}$}
\newcommand{\ep}{\varepsilon}

\date{}
\title{The calculation of 2-loop self-energy diagrams \\
 by the sector decomposition\footnote{%
Kogakuin Daigaku Kenkyuronso, vol.62-2, pp.19-34, 2025,  
https://doi.org/10.57377/0002002705 }} 
\author{Kiyoshi KATO\footnote{e-mail\ :\ {\tt katok@kute.tokyo}} \\
{ \small Kogakuin University, Nishi-Shinjuku 1-24, Shinjuku, Tokyo 160-0023, Japan} 
}
\date{}

\begin{document}
%----------------------------------------------

\maketitle

\begin{abstract}
Detailed description of the calculation of the 2-loop self-energy 
for a scalar particle is presented. 
By employing a simple sector decomposition method, 
the ultraviolet divergent part is efficiently separated from the finite part. 
The resulting expression can be used 
for both analytic and numerical computation to renormalize the divergence 
and to provide finite results for physics.
\end{abstract}

%--------------------------
\section{Introduction}

The importance of radiative corrections is well recognized in high-energy physics.
First of all, to study high-precision data, analyses beyond the leading order are essential. 
Also it may be possible to detect the presence of as yet undiscovered particles which might
appear in a loop diagram.
In the electro-weak theory, various mass parameters appear in the computation, 
making it indispensable to develop methods that deal with multi-loop integrals for massive particles.

The sector decomposition (SD) method is well-established 
for calculating loop \\ integrals.\rref{1,2,3,4,5,6,7}
We have employed the SD method in a simple form and applied it to 3-loop integrals.\rref{8}
Once the formula is ready, we can compute it numerically using non-adaptive 
or adaptive integration with transformations and/or extrapolation.\rref{9,10,11}
In the loop integral, two types of singular behavior appear,
i.e., the ultraviolet (UV) divergence and the singularity related to the threshold.
To manage this, we use a double extrapolation technique and
have already shown some results for the 2-loop and for the 3-loop order.\rref{12,13}

%%%~\cite{nakanishi71,binoth00,binoth04,binoth04a,heinrich08a,heinrich21}
%%~\cite{jocs11,cpc16,ddacat17}

In this paper, detailed formulae are presented for the calculation 
of 2-loop integrals with non-trivial numerators. 
The target process is the 2-loop self-energy diagrams for a scalar particle, 
especially for the study of the Higgs particle.\rref{14,15,16,17,18} 
The formalism can be applied to a fermion or a vector particle 
when appropriate form factors are introduced.

We study the 2-loop integral with $N$-propagators
\begin{equation}
{\bf I}=\int  [d\ell_1]\, [d\ell_2]\,
\frac{\cal{N}}{D_1\ldots D_N}
\label{eq:starteq}
\end{equation}
where $[d\ell]= d^n \ell/(i(2\pi)^n)$, 
 $D_j=p_j^2-m_j^2$ is the inverse propagator of the $j$-th line,
$\ell_1$ and $\ell_2$ are the loop momenta and
$\cal{N}$ is the numerator. 
Here $n$ is the space-time dimension and $n=4-2\ep$.
We use the dimensional regularization where the UV  divergence
appears as the inverse-power of $\ep$.

In the standard model(SM), we have 3- and 4-particle vertices. 
The one-particle irreducible diagrams of a scalar particle 
for self-energy diagrams (S1 $\sim$ S8) and tadpole diagrams (T1 $\sim$ T3) are shown in \zu{twoloop}.
In the figure, the number $j$ indicates that the mass of the propagator is $m_j$ and
the assigned Feynman parameter is $x_j$.
S3a and S3b are the symmetric pair and we only consider S3a, referring to it as S3 hereafter. 
Here, $P$ is the momentum entering the self-energy, 
and we denote $s=P^2$.
Diagrams S6, S7, S8 and T3 are expressed as the product of two 1-loop contributions. 
Since 4-particle vertices in SM are independent of momentum, S5 is treated as S3 at $s=0$. 
Similarly, T1 and T2 can be calculated using S3 and S4 at $s=0$, respectively. 
Thus, in this paper, we mainly concentrate our attention on S1, S2, S3 and S4.

\begin{figure}[htb]
\begin{center}
\includegraphics[width=158.2mm]{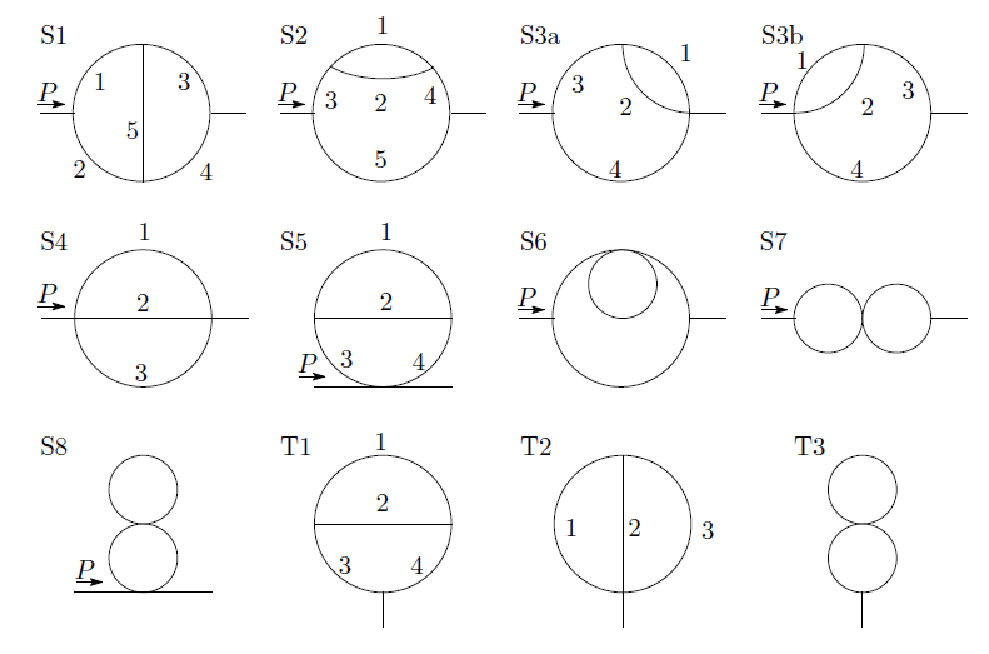}
\end{center}
\caption{One-particle-irreducible 2-loop self-energy diagrams and tadpole diagrams}
\label{fig:twoloop}
\end{figure}

%--------------------------
\section{Scalar integral}

In this section, the components to calculate the loop integrals in \zu{twoloop} are shown and
the strategy to use the SD method is presented.
Here, we study the loop-integral whose numerator ${\cal N}$ is $1$,
and that with non-trivial numerator is treated in the next section.
Based on the Nakanishi-Kinoshita-Cvitanovi\'{c} formula,\rref{19,20} the integral
is given by 
\begin{equation}
{\bf I}=(-1)^N\frac{\Gamma(N-n)}{(4\pi)^{n}}
\int \prod dx_{j} 
\frac{\delta(1-\sum x_j)}{U^{n/2} V^{N-n}} 
= (-1)^N (FAC_2)
\times \frac{\Gamma(N - 4 + 2\ep )}{\Gamma(1+2\ep)}  \times I \, .
\label{eq:integuv}
\end{equation}
\begin{equation}
V=M^2-s\frac{W}{U} , \qquad M^2=\sum_{j=1}^N m_j^2 x_j - i \rho , \qquad
(FAC_2)=\frac{\Gamma(1+2\ep)}{(4\pi)^{4-2\ep}} ,
\end{equation}
\begin{equation}
I=\int d{\bf \Gamma}
\frac{V^{4-N-2\ep}}{U^{2-\ep} }, \qquad
d{\bf \Gamma} = \prod_{j=1}^N dx_{j} \delta(1-\sum x_j)
\label{eq:integzero}
\end{equation}
where $U$ and $W$ functions are given in the following, and
$\rho$ is an infinitesimal positive number. 
The derivation of above formulae is described in the Appendix A.
Here, the notation $x_{jk\ldots}=x_j+x_k+\cdots$ is used.

\begin{equation}
\begin{array}{ll}
\mathrm{S1} & U= x_{12}x_{34}+x_5x_{1234}, \\
  &  W=x_1x_2x_{34}+x_3x_4x_{12}+x_{13}x_{24}x_5 ,\\
\mathrm{S2} & U=x_{12}x_{345}+x_1x_2 , \\
  &  W=x_5(x_{12}x_{34}+x_1x_2) ,\\
\mathrm{S3} & U=x_{12}x_{34}+x_1x_2 , \\
  &  W=x_4(x_1x_2+x_2x_3+x_3x_1) ,\\
\mathrm{S4} & U= x_1x_2+x_2x_3+x_3x_1, \\
  &  W=x_1x_2x_3 .\\
%\mathrm{S5} & U= x_1x_2+x_{12}x_{34}. \\
%  &  W=0 .\\
\end{array}
\end{equation}

We perform the variable transformation. Here $\bar{a}=1-a$.

\begin{equation}
\begin{array}{llllll}
\mathrm{S1} & x_1=y_1\bar{w}_1 & x_2=y_1w_1  & x_3=y_2\bar{w}_2 & x_4=y_2w_2 &
 x_5= y_3 \, ; \\
\mathrm{S2} & x_1=y_1 & x_2=y_2 &
x_3=y_3 \bar{w}\bar{z} & x_4=y_3 \bar{w}z & x_5=y_3 w \, ;\\
\mathrm{S3} & x_1=y_1 & x_2=y_2 &
x_3=y_3\bar{w} & x_4=y_3 w \, ;\\
\mathrm{S4} &  x_1=y_1 & x_2=y_2 & x_3=y_3 .
\end{array}
\end{equation}

After the transformation, $U$ and $W$ are as follows.
The $U$ is common for all S1 $\sim$ S4.
\begin{equation}
\begin{array}{ll}
\mathrm{ALL} & U=y_1y_2+y_1y_3+y_2y_3 \, , \\
\mathrm{S1} & W=\bar{w}_1w_1 y_1 U + \bar{w}_2w_2 y_2 U
+ (w_1-w_2)^2 y_1y_2y_3 , \\
\mathrm{S2} & W=\bar{w}w y_3 U+ w^2 y_1y_2y_3 , \\
\mathrm{S3} & W=\bar{w}w y_3 U+ w^2 y_1y_2y_3 , \\
\mathrm{S4} &  W=  y_1y_2y_3 \,. \\
%%\mathrm{S5} &  W=  0 \, .\\
\end{array}
\label{eq:wfunc}
\end{equation}
We call the form of $U$ is ``complete'' in the sense that all
possible monomials in $y_1, y_2$ and $y_3$ are present.
This property is advantageous for the SD.

The phase space for each diagram is given by

\begin{equation}
d {\bf \Gamma}= \left\{
\begin{array}{ll}
\mathrm{S1} & dy_1 dy_2 dy_3 \delta(1-\sum y_j)\, y_1y_2 dw_1 dw_2 \, , \\
\mathrm{S2} & dy_1 dy_2 dy_3 \delta(1-\sum y_j)\, y_3^2 \bar{w} dw dz \, , \\
\mathrm{S3} & dy_1 dy_2 dy_3 \delta(1-\sum y_j)\, y_3 dw \, , \\
\mathrm{S4} & dy_1 dy_2 dy_3 \delta(1-\sum y_j)\,  .
\end{array} \right.
\end{equation}
It should be noted that if we assume $m_3=m_4$ in S2, $z$ does not appear in the integrand.
The integral in \siki{integzero} can be calculated using the components presented so far.

\vspace{6mm}

We introduce the SD for the improved treatment of the integral.
There are $6(=3!)$ sectors. The label of a sector, '$k, l, m$', is a permutation of $1, 2, 3$.
\begin{equation}
I=\sum I(klm)
\label{eq:integsum}
\end{equation}

In each sector, the variables $s_k$, $t_l$ and $u_m$ are defined as
\begin{equation}
(klm) \qquad y_k=s_k, \quad y_l=s_k t_l, \quad y_m=s_k t_l u_m
\label{eq:sdy}
\end{equation}
where $0 \le t_l \le 1, \ 0 \le u_m \le 1$. Then

\begin{equation}
dy_1 dy_2 dy_3 \delta(1-\sum y_j) = s_k^3 t_l dt_l du_m ,\qquad
s_k^{-1}=1 + t_l + t_l u_m \, .
\end{equation}

We define $f$ and $q$ in the following formulae.
Since $U$ is complete, $f$ is the same in all sectors.
The $q$ in each sector is calculated from $W$ in \siki{wfunc}.
\begin{equation}
U= s_k^2 t_l f ,\quad W= s_k^3 t_l q , \quad
G_{klm}= V/s_k= \tilde{M}^2 - s \frac{q}{f} , \quad \tilde{M}^2=M^2/s_k=\sum_{j=1}^N m_j^2 (x_j/s_k)
\end{equation}
where $x_j/s_k$ in $\tilde{M}^2$ is written by $t_l, u_m$, $w$'s and $z$.

Hereafter, we omit the suffix of $t, u, G$ when it is redundant.

\begin{equation}
f= 1 + u + tu ,\qquad g= s_k^{-1}=1+t+tu \, .
\end{equation}

\begin{equation}
\begin{array}{ll}
\mathrm{S1}    & G=\tilde{M}^2  - s \left( \bar{w}_1w_1 \eta_1 
  + \bar{w}_2w_2 \eta_2 + (w_1-w_2)^2 R \right) , \\
  & \qquad \tilde{M}^2 = m_1^2 \eta_1 \bar{w}_1 + m_2^2 \eta_1 w_1
    +m_3^2 \eta_2 \bar{w}_2 + m_4^2 \eta_2 w_2 + m_5^2 \eta_3 , \\ 
\mathrm{S2} & G=\tilde{M}^2  - s ( \bar{w}w \eta_3 + w^2 R ) , \\
  & \qquad \tilde{M}^2 = m_1^2 \eta_1  + m_2^2 \eta_2 
     + m_3^2 \eta_3 \bar{w}\bar{z} + m_4^2 \eta_3 \bar{w} z + m_5^2 \eta_3 w , \\ 
\mathrm{S3} & G=\tilde{M}^2  - s ( \bar{w}w \eta_3 + w^2 R ) , \\
  & \qquad \tilde{M}^2 = m_1^2 \eta_1  + m_2^2 \eta_2 
     + m_3^2 \eta_3 \bar{w} + m_4^2 \eta_3 w , \\ 
\mathrm{S4} & G=\tilde{M}^2  - s R \, , \\
  & \qquad \tilde{M}^2 = m_1^2 \eta_1  + m_2^2 \eta_2 + m_3^2 \eta_3 \, \\
%%\mathrm{S5} & G=\tilde{M}^2 \, . \\
\end{array}
\end{equation}
where
\begin{equation}
R=  \frac{y_1y_2y_3}{Us_k}=\frac{tu}{f} , 
\end{equation}
and
\begin{equation}
\eta_j= \frac{y_j}{s_k} = 
\left\{ 
  \begin{array}{ll}
  1 & j=k \\
  t & j=l \\
  tu & j=m 
  \end{array}
\right.  \quad {\mathrm{in \ sector\ }} (klm)\, .
\end{equation}

In the above expressions,
sector-dependent quantities are $\tilde{M}^2$ and $\eta_j$.
We obtain the formula to compute the loop integral for these diagrams.

\begin{equation}
\begin{array}{ll}
\mathrm{S1} & \rule{0mm}{8mm} 
I(klm)= {\displaystyle \int_0^1 dw_1 \int_0^1 dw_2 \int_0^1 dt \int_0^1 du\, 
 \frac{h}{t^{1-\ep}} \frac{1}{f^{2-\ep}G^{1+2\ep}} ,
\quad h=\eta_1 \eta_2 }, \\
\mathrm{S2} &  \rule{0mm}{8mm} 
I(klm)= {\displaystyle \int_0^1 \bar{w} dw \int_0^1 dz \int_0^1 dt \int_0^1 du \,
 \frac{h}{t^{1-\ep}} \frac{1}{f^{2-\ep}G^{1+2\ep}} ,
\quad h= \eta_3^2}, \\
\mathrm{S3} &   \rule{0mm}{8mm} 
I(klm)= {\displaystyle \int_0^1  dw  \int_0^1 dt \int_0^1 du  \,
 \frac{h}{t^{1-\ep}} \frac{1}{f^{2-\ep}G^{2\ep}} ,
\quad h= \eta_3 } , \\
\mathrm{S4} &   \rule{0mm}{8mm} 
I(klm)= {\displaystyle  \int_0^1 dt \int_0^1 du \,
  \frac{h}{t^{1-\ep}} \frac{1}{f^{2-\ep}G^{-1+2\ep}} ,
\quad h=1 }. \\
\end{array}
\label{eq:scalarint}
\end{equation}

The UV divergence 
comes both from the Gamma function in \siki{integuv} for $N<5$ and from the
integral by Feynman parameter. 
The UV divergence originating in the integral of \siki{scalarint} is clear   
since the divergent factor in $t$-integral is explicitly shown.
The singular factor is squeezed out from $U$ to leave the non-singular $f$ function
and this is the important property of the SD.

The separation of the UV singularity is done as shown in the Appendix of Ref.8).
Since in S1 $h$ always has a factor $t$, $I(klm)$ has no UV divergence.
We calculate ${\bf I}$ in \siki{integuv} up to finite term, so that the expansion in $\ep$
is to be done :
{
\begin{itemize} \itemsep=-1pt
\item For S1, $O(1)$ term of $I$ is to be calculated.
\item For S2, $O(\ep^{-1})$ and $O(1)$ terms of $I$ are to be calculated.
\item For S3 and S4, $O(\ep^{-1})$, $O(1)$ and $O(\ep)$ terms of $I$ are to be calculated.
\end{itemize} }

The explicit form of the expansion is shown in Appendix D.

When we assume that the scalar is Higgs particle, 
it couples to two particles of same masses if we consider
the gauge in which the mass of Goldstone boson is the same as that of
the corresponding gauge boson.
Then the masses of 
propagators have the following relation.

\begin{equation}
\begin{array}{ll}
\mathrm{S1} & m_1=m_2, \quad m_3=m_4  \\
\mathrm{S2} & m_3=m_4=m_5  \\
\mathrm{S3} & m_3=m_4  \\
%%\mathrm{S5} & m_3=m_4  
\end{array}
\end{equation}
After we rename $m_3, m_5$ as $m_2, m_3$ in S1, $\tilde{M}^2=m_k^2 + m_j^2 t + m_m^2 tu$
in the sector $(klm)$ for all diagrams.
Also in S2, $z$ does not appear in the integrand.

%--------------------------
\section{Integral with the numerator}

In this section, we study the loop-integral whose numerator ${\cal N}$ is
dependent on the loop momenta.
We assume that in the numerator there is no term in which  
the sum of the degree of $\ell_1$ and that of $\ell_2$ is greater than 4.
Though the assignment is arbitrary, one must fix the flow of loop momenta and external
momentum to determine the coefficients for a given diagram. 
In the \zu{flowloop} our standard one is shown. 

\begin{figure}[htb]
\begin{center}
\includegraphics[width=136.4mm]{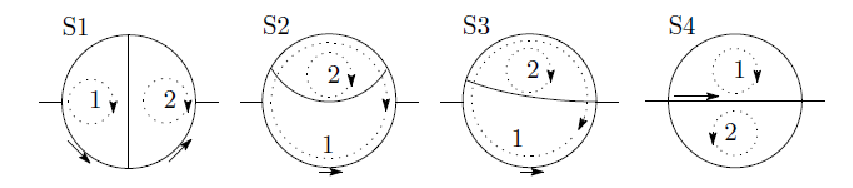}
\end{center}
\caption{Loops for each diagram.
A dotted line shows the flow of the loop momentum, 
with the number 1 or 2 indicating $\ell_1$ or $\ell_2$, respectively. 
A solid arrow shows the flow of the external momentum $P$.
}
\label{fig:flowloop}
\end{figure}

%\begin{center}
%\begin{tabular}{llll}
%\hline
%diagram & $\ell_1$ & $\ell_2$ & $P$ \\
%\hline
%S1 &  $  1 \rightarrow 5  \rightarrow 2   $ & 
%      $  3 \rightarrow 4  \rightarrow 5   $ & 
%      $  2 \rightarrow 4  $ \\ 
%S2 &  $  3 \rightarrow 1 \rightarrow 4 \rightarrow 5  $ & 
%      $  1 \rightarrow 2  $ & 
%      $  5  $ \\ 
%S3 &  $  3 \rightarrow 1 \rightarrow 4  $ & 
%      $  1 \rightarrow 2 $ & 
%      $  4   $ \\ 
%S4 &  $  1 \rightarrow 2 $ & 
%      $  3 \rightarrow 2 $ & 
%      $  2 $ \\ 
%S5 &  $  1 \rightarrow 2   $ & 
%      $  2 \rightarrow 3 \rightarrow 4  $ & 
%      $    $ \\ 
%\hline
%\end{tabular}\end{center}

The numerator is transformed by the process described in Appendix A.
One needs parameters for the transformation and they are as follows.
\begin{equation}
\begin{array}{ll}
\mathrm{S1} &  \alpha=y_{13},\ \beta=y_{23},\ \gamma=-y_3 ,\ b_1=-y_1w_1 , \ b_2=-y_2w_2 \\
\mathrm{S2} &  \alpha=y_{13},\ \beta=y_{12},\ \gamma=y_1 ,\ b_1=-y_3w , \ b_2=0 \\
\mathrm{S3} &  \alpha=y_{13},\ \beta=y_{12},\ \gamma=y_1 ,\ b_1=-y_3w , \ b_2=0 \\
\mathrm{S4} &  \alpha=y_{12},\ \beta=y_{23},\ \gamma=y_2 ,\ b_1=-y_2 , \ b_2=-y_2 \\
%%\mathrm{S5} &  \alpha=y_{12},\ \beta=y_{23},\ \gamma=y_2 ,\ b_1=0 , \ b_2=0 \\
\end{array}
\label{eq:parmlist}
\end{equation}

By \siki{appalast} the formula for the 2-loop integral is 
\[
{\bf I}= (-)^{N} (FAC_2)
 \frac{\Gamma(N-4+2\ep)}{\Gamma(1+2\ep)} \times I 
\]
\[
I= \int d{\bf \Gamma} 
\biggl[
 a_{00} + \frac{-(2-\ep)}{N-5+2\ep}(a_{10}+a_{01}) \frac{V}{U} 
\]
\begin{equation}
 +  \frac{(3-\ep)(2-\ep)}{(N-5+2\ep)(N-6+2\ep)}(a_{20}+\frac{2-\ep}{3-\ep}a_{11}+a_{02}) \frac{V^2}{U^2}
\biggr] \frac{1}{U^{2-\ep}V^{N-4+2\ep}}
\label{eq:integnum}
\end{equation}
where $a_{ij}$'s are expressed by $\alpha, \beta, \gamma, e_1, e_2$.  

In the SM, the 4-particle vertex does not depend on the momentum, so that
$a_{10}=a_{01}=a_{20}=a_{11}=a_{02}=0$ for S4 
and $a_{20}=a_{11}=a_{02}=0$ for S3.  The expression for each diagram is as follows
\begin{equation}
\begin{array}{ll}
\mathrm{S1,S2} & \rule{0mm}{8mm}
I={\displaystyle \int d{\bf \Gamma} 
\left[ {\bf F}_0  + \frac{1}{\ep} {\bf F}_2  \frac{V}{U}
 +  \frac{1}{\ep} {\bf F}_4  \frac{V^2}{U^2}  \right]
 \frac{1}{U^{2-\ep}V^{1+2\ep}} } \, ,
 \\ 
\mathrm{S3} & \rule{0mm}{8mm}
I={\displaystyle \int d{\bf \Gamma} 
\left[  {\bf F}_0  +  {\bf F}_2 \frac{V}{U} \right]
 \frac{1}{U^{2-\ep}V^{2\ep}} } \, ,
 \\ 
\mathrm{S4} & \rule{0mm}{8mm}
I={\displaystyle \int d{\bf \Gamma} 
   {\bf F}_0  \frac{1}{U^{2-\ep}V^{-1+2\ep}} } 
 \\ 
\end{array}
\label{eq:integnuma}
\end{equation}
where
\begin{equation}
\begin{array}{ll}
\mathrm{S1,S2} & \rule{0mm}{8mm}
{\displaystyle {\bf F}_0= a_{00} , \ 
{\bf F}_2= - \left(1-\frac{\ep}{2}\right) (a_{10}+a_{01}) , \ 
{\bf F}_4= - \frac{(3-\ep)(2-\ep)}{2(1-2\ep)} \left(a_{20}+\frac{2-\ep}{3-\ep}a_{11}+a_{02}\right)}, \\
\mathrm{S3} & \rule{0mm}{8mm}
{\displaystyle {\bf F}_0= a_{00} , \quad
{\bf F}_2= \frac{2-\ep}{1-2\ep}( a_{10}+a_{01} ) } , \\
\mathrm{S4} & \rule{0mm}{8mm}
{\bf F}_0= a_{00} \, .
\end{array}
\label{eq:integnumab}
\end{equation}

\vspace{6mm}

Here we introduce the SD using equations
(from \siki{integsum} to \siki{scalarint}) in the last section.
Since 
\begin{equation}
\frac{V}{U}=\frac{G g}{t f} , 
\end{equation}
we have the following formulae for each diagram with numerator.${}^1$
%%\footnote{
%%The region of integrals is to be understood as in \siki{scalarint}.}.

\begin{equation}
\begin{array}{ll}
\mathrm{S1} & \\ 
I(klm)= {\displaystyle \int dw_1 dw_2 dt du 
\left( \frac{1}{t^{1-\ep}} \frac{h {\bf F}_0}{f^{2-\ep}G^{1+2\ep}}
+ \frac{1}{\ep} \frac{1}{t^{2-\ep}} \frac{hg {\bf F}_2}{f^{3-\ep}G^{2\ep}}
+ \frac{1}{\ep} \frac{1}{t^{3-\ep}} \frac{hg^2 {\bf F}_4}{f^{4-\ep}G^{-1+2\ep}}
 \right) }  , \\
\mathrm{S2} & \\
I(klm)= {\displaystyle \int \bar{w} dw dz dt du 
\left( \frac{1}{t^{1-\ep}} \frac{h {\bf F}_0}{f^{2-\ep}G^{1+2\ep}}
+ \frac{1}{\ep} \frac{1}{t^{2-\ep}} \frac{hg {\bf F}_2}{f^{3-\ep}G^{2\ep}}
+ \frac{1}{\ep} \frac{1}{t^{3-\ep}} \frac{hg^2 {\bf F}_4}{f^{4-\ep}G^{-1+2\ep}}
 \right) }  , \\
\mathrm{S3} &  \\
I(klm)= {\displaystyle \int  dw  dt du 
\left( \frac{1}{t^{1-\ep}} \frac{h {\bf F}_0}{f^{2-\ep}G^{2\ep}}
+ \frac{1}{t^{2-\ep}} \frac{hg {\bf F}_2}{f^{3-\ep}G^{-1+2\ep}}
\right) } , \\
\mathrm{S4} &  \\
I(klm)= {\displaystyle  \int dt du 
\frac{1}{t^{1-\ep}} \frac{{\bf F}_0}{f^{2-\ep}G^{-1+2\ep}} }. \\
\end{array}
\label{eq:integsdnum}
\end{equation}

In \siki{integsdnum}, the UV-divergent factor is generated by the $t$-integral 
whose generic form is
\begin{equation}
J_r= \int_0^1 dt \frac{Q(t)}{t^{r-\ep}} \qquad (r=0,1,2,3) 
\label{eq:integjn}
\end{equation}
where $Q(t)$ is a function which is differentiable by $t$ and finite at $t=0$.

We make inspection how many $t$ appears in the numerator of \siki{integsdnum}. 
As is already described,
the ${\bf F}'s$ are expressed by $\alpha, \beta, \gamma, e_1, e_2$ in \siki{parmlist}.

The $e_1$ and $e_2$ have no contribution to the factor $t$. They are given in
Appendix A as 
\begin{equation}
e_1= \frac{(gb_1)(g\beta)-(gb_2)(g\gamma)}{t f }, \quad
e_2= \frac{-(gb_1)(g\gamma)+(gb_2)(g\alpha)}{t f} .
\end{equation}
When we use the expression in \siki{parmlist}, except for the S4 which needs no $e_1$ nor $e_2$,
 $t$ in the denominator is canceled by the same factor in the numerator,
so that $e_1$ and $e_2$ are regular at $t=0$.

The numerator structure is studied in Appendix B.
{
\begin{itemize} \itemsep=-1pt
\item ${\bf F}_0$ has no $\alpha, \beta, \gamma$.
\item ${\bf F}_2$ % has 1st order dependence on $\alpha, \beta, \gamma$, i.e.,  ${\bf F}_2$ 
is expanded by $\alpha, \beta, \gamma$.
\item ${\bf F}_4$ %has 2nd order dependence on $\alpha, \beta, \gamma$, i.e., ${\bf F}_4$ 
is expanded by $\alpha^2, \beta^2, \gamma^2, \alpha \beta, \alpha\gamma, \beta\gamma $.
\end{itemize} }

Then $g{\bf F}_2$ and $g^2{\bf F}_4$ are just ${\bf F}_2$ and 
${\bf F}_4$ after replacing $\alpha, \beta, \gamma$ by $g\alpha, g\beta, g\gamma$.
The latter, $g\alpha, g\beta, g\gamma$, are $\alpha, \beta, \gamma$ in \siki{parmlist}
after replacing $y_j$ by $\eta_j$.

For each diagram, the factor $t$ is counted using \siki{parmlist} as follows.
We use the notation $A \sim t^r$ when
 $r$ is the minimum number of factor $t$ in $A$ among all sectors.

{
\begin{itemize} \itemsep=-1pt
\item 
S1 \\
In this diagram, $h\sim t^1$.
And $h\alpha, h\beta \sim t^1$, $h\gamma \sim t^2$,
$h\alpha^2, h\beta^2 \sim t^1$, $h\gamma^2,h\alpha \beta, h\alpha\gamma, h\beta\gamma \sim t^2$.
Since $\ell_1$ and $\ell_2$ flow along semicircles in the diagram,
neither $f_{16}$(has $\beta^2$) nor $f_{18}$(has $\alpha^2$) appears in ${\bf F}_4$.
Then the calculation of $I(klm)$ needs $J_0$- and $J_1$-type integral.

\item S2 \\
In this diagram, $h\sim t^0$.
And $h\alpha \sim t^0$, $h\beta, h\gamma \sim t^1$,
$h\alpha^2 \sim t^0$, $ h\alpha \beta, h\alpha\gamma \sim t^1$,
$h\beta^2, h\gamma^2, h\beta\gamma \sim t^2$.
Since $\ell_2$ flows in the small loop, 
$f_{18}$(has $\alpha^2$) does not appear in ${\bf F}_4$.
Then the calculation of $I(klm)$ needs $J_0$-, $J_1$- and $J_2$-type integral.

\item S3\\
In this diagram, $h\sim t^0$ and no ${\bf F}_4$.
And $h\alpha \sim t^0$, $h\beta, h\gamma \sim t^1$.
Then the calculation of $I(klm)$ needs $J_0$-, $J_1$- and $J_2$-type integral.

\item S4 \\
There is only ${\bf F}_0$.
Then the calculation of $I(klm)$ needs $J_1$-type integral.

%\item S5 \\
%In this diagram, $h\sim t^0$ and no ${\bf F}_4$.
%And $h\beta \sim t^0$, $h\alpha, h\gamma \sim t^1$.
%Then the calculation of $I(klm)$ needs $J_0$-, $J_1$-and $J_2$- type integral.
\end{itemize} }

As is shown, $J_3$-type integral does not show up.
When one computes \siki{integsdnum}, in each sector one first checks 
the type of integral by counting factor $t$ in the numerator,
and performs the integration. When it is $J_0$,
there is no singular behavior at $t=0$ and it can be integrated as it is. 
The procedure to integrate $J_1$ and $J_2$ is shown in Appendix C.

%where
%\begin{equation}
%{\bf F}_0= a_{00},\quad
%{\bf F}_2= \frac{-(2-\ep)}{N-5+2\ep}(a_{10}+a_{01}),\quad
%{\bf F}_4= \frac{(3-\ep)(2-\ep)}{(N-5+2\ep)(N-6+2\ep)}(a_{20}+\frac{2-\ep}{3-\ep}a_{11}+a_{02}) 
%\end{equation}

%--------------------------
\section{Summary}

In this paper, we have presented the detailed formulae 
which enable the calculation of the 2-loop self-energy for a scalar particle. 
By using the SD method, the ultraviolet divergence is under control. 
The results can be expressed as a series in $\ep$. 
While a few leading terms of the series can be obtained analytically, 
complete computation must be done numerically.

In the numerical computation, there are two approaches.
One way is to expand the formula symbolically in $\ep$ and integrate 
each term numerically. Another way is to compute the formula, 
including $\ep$, numerically for several values of $\ep$, 
and to estimate the coefficients of the $\ep$-series by extrapolation or using a linear solver. 
Both methods have been tested and have shown good results in our preceding research.\rref{8,12,13}

The work presented here shall be useful for the further study of
the electro-weak theory as the radiative correction in the 2-loop level.

\vspace{7mm}

\noindent{\itshape Acknowledgement}\ 
The author acknowledges Dr. F. Yuasa and Dr. T. Ishikawa 
%for their assistance 
%in verifying the analytic results through detailed numerical computations, 
%as well as 
for their helpful discussions and encouragement. 
The author also thanks Dr. T. Kaneko for explaining the SD method.

%--------------------------
%%\appendix
\begin{appendices}
\section{Derivation of the integral formula}

In this appendix, we briefly describe the procedure to
obtain \siki{integuv} and \siki{integnum} from \siki{starteq}.
While the procedure is applicable to general multi-loop integral,
we present the case of 2-loop integral.
Assumptions in the last sections are kept in this Appendix.

\begin{equation}
{\bf I}=\int [d\ell_1]\,[d\ell_2]\,
 \frac{{\cal N}}{(p_1^2-m_1^2)\cdots(p_N^2-m_N^2)}
= \int d{\bf \Gamma} \Gamma(N) \int  [d\ell_1]\,[d\ell_2]\,
\frac{{\cal N}}{{\cal D}^N}
\end{equation}

\begin{equation}
{\cal D}=\sum x_r (p_r^2-m_r^2) 
= {}^t \vec{\ell}{\cal A} \vec{\ell} + 2\,  {}^t \vec{\ell} \ \vec{b} + C
\end{equation}
where
\begin{equation}
\vec{\ell}=\left(
\begin{array}{c} \ell_1  \\ \ell_2 \end{array}
\right),\quad
{\cal A}=\left(
\begin{array}{cc} \alpha & \gamma \\ \gamma & \beta \end{array}
\right),\quad
\vec{b}=\left(
\begin{array}{c} b_1  \\ b_2 \end{array}
\right),\quad
{\cal C}=\left(
\begin{array}{cc} {\cal A} & \vec{b} \\ {}^t\vec{b} & C \end{array}
\right) 
\end{equation}
and $U=\det {\cal A} ,\ -UV=\det{\cal C}$.

We perform the three steps of the variable transformation to
make ${\cal D}$ into the form of $\bar{\ell}_1^2+\bar{\ell}_2^2-V$.

\vspace{3mm}
\noindent Step-1 Shift \\
\begin{equation}
\vec{\ell}=\vec{\ell'}-{\cal A}^{-1}\vec{b} \cdots
\left\{ \begin{array}{l}
\ell_1 = \ell'_1 - e_1 P \\
\ell_2 = \ell'_2 - e_2 P,
\end{array} \right. ,\quad 
e_1= \frac{b_1\beta-b_2\gamma}{U}, \quad
e_2= \frac{-b_1\gamma+b_2\alpha}{U}.
\end{equation}
By this, the second term of ${\cal D}$ is erased.
%%After this step, the terms are discarded in the numerator such that
%%the sum of the degree of $\ell'_1$ and $\ell'_2$ is odd.

\vspace{3mm}
\noindent Step-2 Rotation \\
\begin{equation}
\vec \ell' = {\cal O} \vec \ell" \cdots
\left\{\begin{array}{c} \ell'_1= C\ell"_1 -S\ell"_2 \\
 \ell'_2= S\ell"_1 +C\ell"_2
 \end{array} \right.  ,\quad
{\cal O}=\left(
\begin{array}{cc} C & -S \\ S & C \end{array}
\right),\quad
C=\cos \theta, S=\sin \theta .\quad
\end{equation}
By this, ${\cal D}$ is made to be diagonal:
\begin{equation}
{}^t {\cal O}{\cal A}{\cal O}=
 \left(
\begin{array}{cc} \lambda_1 & 0 \\ 0 & \lambda_2 \end{array}
\right) .
\end{equation}

After this step, the terms in the numerator are discarded such that
the degree of $\ell"_1$ is odd and/or the degree of $\ell"_2$ is odd.
Also, the following contraction is performed.

\begin{equation}
\ell"_j^{\mu}\ell"_j^{\nu} \rightarrow \ell"_j^2 \frac{1}{n} g^{\mu\nu} .
\end{equation}
\begin{equation}
\ell"_j^{\mu}\ell"_j^{\nu} \ell"_j^{\rho}\ell"_j^{\sigma} \rightarrow 
(\ell"_j^2)^2 \frac{1}{n(n+2)} \left( g^{\mu\nu}g^{\rho\sigma}  
 + g^{\mu\rho}g^{\nu\sigma} + g^{\mu\sigma}g^{\nu\rho} \right) .
\end{equation}

The eigenvalues have the relation $\lambda_1 \ge \lambda_2$.
The following formulae for eigenvalues are useful:
\begin{equation}
U=\det {\cal A} = \lambda_1 \lambda_2 = \alpha\beta - \gamma^2 ,\quad
\lambda_{1,2}=\frac{1}{2}(\alpha+\beta\pm \delta), \quad \delta=\sqrt{(\alpha-\beta)^2+4\gamma^2}.
\end{equation}

\vspace{3mm}
\noindent Step-3 Scaling
\begin{equation}
\ell"_1^2 = \frac{\bar{\ell}_1^2}{\lambda_1} , \quad
\ell"_2^2 = \frac{\bar{\ell}_2^2}{\lambda_2} .
\end{equation}

Final formula is given by

\begin{equation}
{\bf I}= \int d{\bf \Gamma} \Gamma(N) 
\int [d\bar{\ell}_1]\,[d\bar{\ell}_2]\,
\frac{{\cal N}}{(\lambda_1\lambda_2)^{n/2} (\bar{\ell}_1^2+\bar{\ell}_2^2 - V)^N} \,.
\end{equation}

The numerator ${\cal N}$ is processed by the variable transformation,
the removal of odd $\ell$ terms and the contraction of $\ell$'s, so that 
it becomes the following form
\begin{equation}
{\cal N}=a_{00}
  + \frac{a_{10}\,\bar{\ell}_1^2
         +a_{01}\,\bar{\ell}_2^2}{U} 
  + \frac{a_{20}\,(\bar{\ell}_1^2)^2
         +a_{11}\,\bar{\ell}_1^2\,\bar{\ell}_2^2
         +a_{02}\,(\bar{\ell}_2^2)^2}{U^2} , 
\end{equation}
where $a$'s are coefficients. 
As is shown, the factor $U$ appears except for the first term.

In the coefficients $a_{km}$, $C$ or $S$ does not appear 
when we apply the following formulae.

\begin{equation}
CS(\alpha-\beta)=(C^2-S^2)\gamma.
\end{equation}

\begin{equation}
\frac{C^2}{\lambda_1} + \frac{S^2}{\lambda_2}=\frac{\beta}{\lambda_1\lambda_2},\quad
\frac{S^2}{\lambda_1} + \frac{C^2}{\lambda_2}=\frac{\alpha}{\lambda_1\lambda_2},\quad
CS\left(\frac{1}{\lambda_1} - \frac{1}{\lambda_2}\right)=\frac{-\gamma}{\lambda_1\lambda_2}.
\end{equation}

The integration by $\bar{\ell}_1$ and $\bar{\ell}_2$ results 
\begin{equation}
{\bf I}= \int d{\bf \Gamma} 
\sum a_{km} \, (-)^{N+k+m}\frac{\Gamma(N-n-k-m)}{(4\pi)^n}
\frac{\Gamma(n/2+k)}{\Gamma(n/2)} 
\frac{\Gamma(n/2+m)}{\Gamma(n/2)} 
\frac{1}{U^{n/2+k+m}V^{N-n-k-m}} \, .
\label{eq:appalast}
\end{equation}

%--------------------------
%%%\appendix
\section{Decomposition of the numerator}

The explicit representation of ${\bf F}_0$, ${\bf F}_2$ and ${\bf F}_4$ can be achieved 
through the application of the sequence of transformations given in Appendix A to ${\cal N}$. 
Normally, this is done automatically by symbolic manipulation software. 
In this appendix, 
we present formulae that are useful for the manual analysis of the numerator structure.

The numerator has the following form.

\begin{equation}
{\cal N}= \sum_{j=1}^{21} C_j f_j 
\end{equation}
where the coefficient $C_j$ includes coupling constants,
masses, external momenta and $n$ but $x_j$.
We choose $f_j$ functions as follows.

\begin{equation}
\begin{array}{llllll}
f_1= 1, & f_2= (\ell_1P), & f_3= (\ell_2P), \\ 
f_4= (\ell_1^2), & f_5= (\ell_1\ell_2), & f_6=(\ell_2^2) , \\ 
f_7= (\ell_1P)^2, & f_8=(\ell_1P)(\ell_2P) , & f_9= (\ell_2P)^2, \\ 
f_{10}=(\ell_1^2)(\ell_1P) , & f_{11}=(\ell_1\ell_2)(\ell_1P) , & f_{12}=(\ell_2^2)(\ell_1P) , \\ 
f_{13}=(\ell_1^2)(\ell_2P) , & f_{14}=(\ell_1\ell_2)(\ell_2P) , & f_{15}=(\ell_2^2)(\ell_2P) , \\
f_{16}=(\ell_1^2)^2 , & f_{17}=(\ell_1\ell_2)^2 , & f_{18}=(\ell_2^2)^2 , \\
f_{19}=(\ell_1^2)(\ell_1\ell_2) , & f_{20}=(\ell_2^2)(\ell_1\ell_2) , & f_{21}=(\ell_1^2)(\ell_2^2)  .  
\end{array}
\end{equation}

After the transformation given in Appendix A, $f_j$ becomes $\tilde{f}_j$
which is expanded by $[k,m]=(\bar{\ell}_1^2)^k (\bar{\ell}_2^2)^m$.
When we take the loop-momentum integral into account the following identification holds
\begin{equation}
[1,0]=[0,1], \qquad [2,0]=[0,2], \qquad [1,1]=\frac{n}{n+2}[2,0] ,
\end{equation}
so that $\tilde{f}_j$ can be expressed as 
\begin{equation}
\tilde{f}_j= \tilde{f}_j^{(0)} [0,0]  +  \frac{1}{U} \tilde{f}_j^{(1)} [1,0] 
   +  \frac{1}{U^2} \tilde{f}_j^{(2)} [2,0] \, 
\end{equation}
and $\tilde{f}_j^{(0)}=a_{00}$, $\tilde{f}_j^{(1)}=a_{10}+a_{01}$, 
$\tilde{f}_j^{(2)}=a_{20}+(\ep-2)/(\ep-3)a_{11}+a_{02}$.
The $\tilde{f}_j^{(k)}$ is expressed by $\alpha, \beta, \gamma, e_1, e_2$.
Explicit expressions of $\tilde{f}_j$ are presented below:${}^2$
%%\footnote{
%%The first term of $f_{17}$ and that of $f_{21}$ are
%%\[
%%\tilde{f}_{17}= \frac{\gamma^2}{U^2}[2,0]+\frac{1}{n}\frac{1}{U}[1,1]+\cdots,\qquad
%%\tilde{f}_{21}= \frac{\gamma^2}{U^2}[2,0]+\frac{1}{U}[1,1] +\cdots .
%%\]
%%}

\begin{equation}
\tilde{f}_1= 1\cdot [0,0]
%\end{equation}
,\quad
%\begin{equation}
\tilde{f}_2= -e_1 s  [0,0]
%\end{equation}
,\quad
%\begin{equation}
\tilde{f}_3=  -e_2 s  [0,0] ,
\end{equation}

\begin{equation}
\tilde{f}_4= \frac{\beta}{U}[1,0]+e_1^2 s [0,0]
%\end{equation}
,\quad
%\begin{equation}
\tilde{f}_5= \frac{-\gamma}{U}[1,0]+e_1 e_2 s [0,0]
%\end{equation}
,\quad
%\begin{equation}
\tilde{f}_6= \frac{\alpha}{U}[1,0]+e_2^2 s [0,0] ,
\end{equation}

\begin{equation}
\tilde{f}_7= \frac{s}{n} \frac{\beta}{U}[1,0]+e_1^2 s^2 [0,0]
%\end{equation}
,\quad
%\begin{equation}
\tilde{f}_8= \frac{s}{n} \frac{-\gamma}{U}[1,0]+e_1e_2 s^2 [0,0]
%\end{equation}
,\quad
%\begin{equation}
\tilde{f}_9= \frac{s}{n} \frac{\alpha}{U}[1,0]+e_2^2 s^2 [0,0] ,
\end{equation}

\begin{equation}
\tilde{f}_{10}= -e_1 s (1+\frac{2}{n}) \frac{\beta}{U}[1,0]  - e_1^3 s^2 [0,0]
\end{equation}

\begin{equation}
\tilde{f}_{11}= -\left( e_1 s (1+\frac{1}{n}) \frac{-\gamma}{U} 
 +e_2 \frac{s}{n} \frac{\beta}{U}\right)[1,0]
  - e_1^2e_2 s^2 [0,0]
\end{equation}

\begin{equation}
\tilde{f}_{12}= - \left( e_1 s \frac{\alpha}{U} + 2e_2 \frac{s}{n}\frac{-\gamma}{U})\right) [1,0]  
- e_1e_2^2 s^2 [0,0]
\end{equation}

\begin{equation}
\tilde{f}_{13}= - \left(e_2 s \frac{\beta}{U}+ 2e_1\frac{s}{n}\frac{-\gamma}{U}) \right) [1,0]
  - e_1^2 e_2 s^2 [0,0]
\end{equation}

\begin{equation}
\tilde{f}_{14}= - \left(e_2 s (1+\frac{1}{n}) \frac{-\gamma}{U}
+ e_1\frac{s}{n}\frac{\alpha}{U}) \right) [1,0]
  - e_1 e_2^2 s^2 [0,0]
\end{equation}

\begin{equation}
\tilde{f}_{15}= - e_2 s (1+\frac{2}{n}) \frac{\alpha}{U}[1,0]  - e_2^3 s^2 [0,0]
\end{equation}

\begin{equation}
\tilde{f}_{16}= \frac{\beta^2}{U^2}[2,0]
 + 2 e_1^2 s (1+\frac{2}{n}) \frac{\beta}{U}[1,0]  
 + e_1^4 s^2 [0,0]
\end{equation}

\begin{equation}
\tilde{f}_{17}= \frac{1}{n+2} \frac{(n+1)\gamma^2+\alpha\beta}{U^2}[2,0]
 + \left\{  e_1^2 \frac{s}{n} \frac{\alpha}{U} + e_2^2 \frac{s}{n}\frac{\beta}{U} 
 + 2e_1e_2 s(1+\frac{1}{n}) \frac{-\gamma}{U} 
 \right\} [1,0]
 + e_1^2 e_2^2 s^2 [0,0]
\end{equation}

\begin{equation}
\tilde{f}_{18}= \frac{\alpha^2}{U^2}[2,0]
 + 2 e_2^2 s (1+\frac{2}{n}) \frac{\alpha}{U}[1,0]  
 + e_2^4 s^2 [0,0]
\end{equation}

\begin{equation}
\tilde{f}_{19}= \frac{-\beta\gamma}{U^2}[2,0]
 + \left\{ 2 e_1^2 \frac{s}{n} \frac{-\gamma}{U} + 2 e_1e_2 \frac{s}{n} \frac{\beta}{U}
  +  e_1^2 s \frac{-\gamma}{U} +  e_1e_2 s \frac{\beta}{U}  \right\} [1,0]  
 + e_1^3 e_2 s^2 [0,0]
\end{equation}

\begin{equation}
\tilde{f}_{20}= \frac{-\alpha\gamma}{U^2}[2,0]
 + \left\{ 2 e_2^2 \frac{s}{n} \frac{-\gamma}{U} + 2 e_1e_2 \frac{s}{n} \frac{\alpha}{U}
  +  e_2^2 s \frac{-\gamma}{U} +  e_1e_2 s \frac{\alpha}{U}  \right\} [1,0]  
 + e_1 e_2^3 s^2 [0,0]
\end{equation}

\begin{equation}
\tilde{f}_{21}= \frac{1}{n+2} \frac{2\gamma^2+n\alpha\beta}{U^2}[2,0]
 + \left\{  e_1^2 s \frac{\alpha}{U} + e_2^2 s \frac{\beta}{U} 
 + 4 e_1e_2 \frac{s}{n} \frac{-\gamma}{U} 
 \right\} [1,0]
 + e_1^2 e_2^2 s^2 [0,0]
\end{equation}

%--------------------------
%%%\appendix
\section{Singular integrals}

There are several possibility to deal with $J_r$ in \siki{integjn}.
One needs the analytic continuation for the evaluation at $t=0$.
In the Appendix of Ref.8), two methods, the expansion and the integration
by parts, are given. 
Here we present several forms based on the integration by parts.
The choice of which of these expressions to use is 
arbitrary and depends on the situation.

\begin{equation}
J_1=\frac{1}{\ep}\left( Q(1)-\int_0^1 dt\, t^{\ep} \frac{dQ}{dt} \right)
\end{equation}

\begin{equation}
J_1=\frac{1}{\ep}Q(0)-\int_0^1 dt\, E_t \frac{dQ}{dt}
\end{equation}

\begin{equation}
J_1=\frac{1}{\ep}Q(0)+\int_0^1 dt\, \frac{d E_t}{dt} ( Q-Q(0) )
\label{eq:appcthr}
\end{equation}

\begin{equation}
J_2=\frac{1}{1-\ep} \left[ \frac{1}{\ep}
 \left(\left.\frac{d Q}{dt}\right|_{t=1}  -\int_0^1 dt\, t^{\ep} \frac{d^2Q}{dt^2} \right)
 -Q(1) \right]
\end{equation}

\begin{equation}
J_2= \frac{1}{1-\ep} \left[ \frac{1}{\ep}\left.\frac{d Q}{dt}\right|_{t=0} 
- Q(1) - \int_0^1 dt\, E_t \frac{d^2Q}{dt^2} \right]
\end{equation}

\begin{equation}
J_2= \frac{1}{1-\ep} \left[ \frac{1}{\ep}\left.\frac{d Q}{dt}\right|_{t=0} 
- Q(1) 
+ \int_0^1 dt\, \frac{d E_t}{dt}\left( \frac{dQ}{dt} - \left.\frac{d Q}{dt}\right|_{t=0}  \right) 
\right]
\end{equation}

Here
\begin{equation}
E_t= \frac{t^{\ep}-1}{\ep}=\log t + \frac{\ep}{2}(\log t)^2 + \frac{\ep^2}{6}(\log t)^3 + \cdots .
\end{equation}

%--------------------------
%%%\appendix
\section{Expressions for constant-numerator case}

In this Appendix, the analytic results for the UV-divergent part
in the case of ${\cal N}=1$ are presented. 
Similar formulae are found in Ref.15). 
Here \siki{appcthr} is applied for the singular integral.

\vspace{3mm}

\noindent $\bullet$ \ S1 \par

\begin{equation}
I= \sum_{klm}
\int_0^1 dw_1\int_0^1 dw_2 \int_0^1 dt\int_0^1 du\, \frac{(h/t)}{f^2G_{klm}} + O(\ep)
\end{equation}
and $h/t=1$ for $klm=123,213$, $h/t=u$ for $klm=132,231$, and  $h/t=tu$ for $klm=312,321$.
The $O(1)$ term is computed numerically.

\vspace{3mm}

\noindent $\bullet$ \ S2 \par

\begin{equation}
I=\frac{1}{\ep} J_{{\mathrm S2}}+ I_{{\mathrm S2}}+ O(\ep) ,
\end{equation}

\begin{equation}
J_{{\mathrm S2}} = \int_0^1dw \int_0^1 dz \frac{\bar{w}}{G_0} , \qquad
G_0= m_3^2 \bar{w}\bar{z} + m_4^2 \bar{w} z + m_5^2 w - s \bar{w}w .
\end{equation}

\[
I_{{\mathrm S2}}= 
 2 \int_0^1 \bar{w} dw \int_0^1 dz \int_0^1 du 
\frac{1}{f_0^2 G_0}(\log f_0 - 2\log G_0)
\]
\[
+ \sum_{klm=312,321} \int_0^1 \bar{w} dw \int_0^1 dz \int_0^1 du \int_0^1 dt
\frac{1}{t}\left( \frac{1}{f^2 G_{klm}} - \frac{1}{f_0^2 G_0} \right)
\]
\begin{equation}
+ \int_0^1 \bar{w} dw \int_0^1 dz \int_0^1 du \int_0^1 dt \left(
 \sum_{klm=132,231} \frac{t}{f^2 G_{klm}} + \sum_{klm=123,213} \frac{tu^2}{f^2 G_{klm}} \right)
\end{equation}
where
$f_0=1+u$.

The $O(1/\ep)$ term, $J_{{\mathrm S2}}$, is elementary${}^3$ 
%%\footnote{
%%If $m_3=m_4=m_5=m$, $J_{{\mathrm S2}} = (2/sv) \arctan (1/v)$ for $s<4m^2$ and
%%$J_{{\mathrm S2}} = (1/sv)[ \log ((1-v)/(1+v)) + i\pi ] $ for $s>4m^2$ where
%%$v= \sqrt{|1-4m^2/s|}$.
%%} 
and the $O(1)$ term, $I_{{\mathrm S2}}$, is computed numerically.

\vspace{3mm}

\noindent $\bullet$ \ S3 \par

\begin{equation}
I=\frac{1}{\ep}+ \left( 1 - 2 J_{{\mathrm S3}}  \right) + \ep I_{{\mathrm S3}}+ O(\ep^2) ,
\end{equation}

\begin{equation}
J_{{\mathrm S3}} = \int_0^1 dw \, \log G_0 , \qquad
G_0= m_3^2 \bar{w} +  m_4^2 w - s \bar{w}w .
\end{equation}

\[
I_{{\mathrm S3}}=
  \int_0^1 dw \int_0^1 du 
\frac{1}{f_0^2}(\log f_0 - 2\log G_0)^2
\]
\[
+  \sum_{klm= \atop 312,321} 
\int_0^1 dw  \int_0^1 du \int_0^1 dt \frac{1}{t} \left[
  \frac{1}{f^2}(\log t + \log f - 2\log G_{klm}) 
                    - \frac{1}{f_0^2}(\log t +\log f_0 - 2\log G_0) 
  \right]
\]
\begin{equation}
+ \int_0^1 dw \int_0^1 du \int_0^1 dt \left(
  \frac{2(1+u)}{f^2} (\log t + \log f)
  -2 \sum_{klm= \atop 132,231} \frac{1}{f^2} \log G_{klm} 
  -2 \sum_{klm= \atop 123,213} \frac{u}{f^2} \log G_{klm} 
  \right)
\end{equation}
where
$f_0=1+u$.

The $O(1)$ term, $J_{{\mathrm S3}}$, is elementary${}^4$ 
%%\footnote{
%%If $m_3=m_4=m$, $ J_{{\mathrm S3}} = \log m^2 -2 + 2v \arctan(1/v)$ for $s<4m^2$ and
%%$ J_{{\mathrm S3}} = \log m^2 -2 + v \log ((1+v)/(1-v)) - i\pi v $ for $s>4m^2$ 
%%where $v$ is the same as in the last footnote.
%%} 
and the $O(\ep)$ term, $I_{{\mathrm S3}}$, is computed numerically.

\vspace{3mm}

\noindent $\bullet$ \ S4 \par

\begin{equation}
I= \frac{1}{\ep} \sum_{j=1}^3 m_j^2 
 + \left( \sum_{j=1}^3 m_j^2 - 2 \sum_{j=1}^3 m_j^2 \log m_j^2 - \frac{1}{2} s \right)
 + \ep I_{{\mathrm S4}} + O(\ep^2)
\end{equation}

\[
I_{{\mathrm S4}} = \sum_{k=1}^3 \int_0^1 du  \frac{G_0}{f_0^2}(\log f_0 - 2\log G_0)^2
\]
\begin{equation}
  + \sum_{klm} \int_0^1 du \int_0^1 dt 
   \frac{1}{t}\left[
  \frac{G_{klm}}{f^2}(\log t + \log f - 2\log G_{klm}) 
                    - \frac{G_0}{f_0^2}(\log t +\log f_0 - 2\log G_0) 
  \right]
\end{equation}
where $f_0=1+u$, $G_0=m_k^2$  and $\sum_{klm}$ is the summation for all $klm$.
The $O(\ep)$ term, $I_{{\mathrm S4}}$, is computed numerically.

\end{appendices}

%--------------------------

\vspace{10mm}

\noindent{\bf \large Notes}

{
\begin{enumerate} \itemsep=0pt
\item
The region of integrals is to be understood as in \siki{scalarint}.

\item
The first term of $f_{17}$ and that of $f_{21}$ are
\[
\tilde{f}_{17}= \frac{\gamma^2}{U^2}[2,0]+\frac{1}{n}\frac{1}{U}[1,1]+\cdots,\qquad
\tilde{f}_{21}= \frac{\gamma^2}{U^2}[2,0]+\frac{1}{U}[1,1] +\cdots .
\]

\item
If $m_3=m_4=m_5=m$, $J_{{\mathrm S2}} = (2/sv) \arctan (1/v)$ for $s<4m^2$ and
$J_{{\mathrm S2}} = (1/sv)[ \log ((1-v)/(1+v)) + i\pi ] $ for $s>4m^2$ where
$v= \sqrt{|1-4m^2/s|}$.

\item
If $m_3=m_4=m$, $ J_{{\mathrm S3}} = \log m^2 -2 + 2v \arctan(1/v)$ for $s<4m^2$ and
$ J_{{\mathrm S3}} = \log m^2 -2 + v \log ((1+v)/(1-v)) - i\pi v $ for $s>4m^2$ 
where $v$ is the same as in the last note.

\end{enumerate}
}

%--------------------------

\vspace{10mm}

%%\newpage
\noindent{\bfseries \large{References}}

\vspace{3mm}

\noindent
1) N.~Nakanishi,
 {\em Graph Theory and {F}eynman Integrals},
  Gordon and Breach, New York, (1971).

\noindent
2) T.~Binoth and G.~Heinrich, Nucl. Phys.~B {\bf 585}, (2000), 741,\par
 DOI: 10.1016/S0550-3213(00)00429-6 .

\noindent
3) T.~Binoth and G.~Heinrich, Nucl. Phys.~B\ {\bf 680}, (2004),  375,\par
  DOI: 10.1016/j.nuclphysb.2003.12.023.

\noindent
4) T.~Binoth and G.~Heinrich, Nucl. Phys.~B, {\bf 693},  (2004), 134, \par
  DOI: 10.1016/j.nuclphysb.2004.06.005.

\noindent
5) G.~Heinrich, Int. J. Mod. Phys.~A  {\bf 23},(2008), 1457, \par
  DOI: 10.1142/S0217751X08040263.

\noindent
6) G.~Heinrich, Physics Reports  {\bf 922},(2021), 1, 
  DOI: 10.1016/j.physrep.2021.03.006.

\noindent
7) T.~Kaneko and T.~Ueda, Comput. Phys. Commun. {\bf 181}, (2010),  1352, \par
  DOI: 10.1016/j.cpc.2010.04.001.

\noindent
8) Elise~de~Doncker, Tadashi~Ishikawa, Kiyoshi~Kato, and Fukuko~Yuasa, \par
Prog. Theor. Exp. Phys. (2024) 083B08 (13 pages), DOI: 10.1093/ptep/ptae122.

\noindent
9) E.~de~Doncker, J.~Fujimoto, N.~Hamaguchi, T.~Ishikawa, Y.~Kurihara, Y.~Shimizu, \par
  and F.~Yuasa, Journal of Computational Science (JoCS) {\bf 3}(3), (2011),  102--112,\par
  DOI: 10.1016/j.jocs.2011.06.003.

\noindent
10) E.~de~Doncker, F.~Yuasa, K.~Kato, T.~Ishikawa, J.~Kapenga, and O.~Olagbemi, \par
  Comput. Phys. Commun. {\bf 224},  (2018),164--185,
  DOI: 10.1016/j.cpc.2017.11.001.

\noindent
11) E.~de~Doncker, A.~Almulihi, and F.~Yuasa, The Journal of Physics: Conf. Ser. (2018), \par
  {\bf 1085}(052005), DOI: 10.1088/1742-6596/1085/5/052005.

\noindent
12) E.~de~Doncker, F.~Yuasa, T.~Ishikawa, and K.~Kato (2022),
To be published in \par
the Proceedings of ACAT2022(Journal of Physics, Conference Series).

\noindent
13) E.~de~Doncker, F.~Yuasa, T.~Ishikawa, and K.~Kato (2023),
To be published in  \par
the Proceedings of CCP2023.

\noindent
14) A.~Ghinculov and J.~J.~van der Bij,
Nucl. Phys. B {\bf 436} (1995), pp.30-48.

\noindent
15) Stephen~P.~Martin,
Phys. Rev. D {\bf 68}, 075002 (2003),
DOI: 10.1103/PhysRevD.68.075002.

\noindent
16) S.~Actis, A.~~Ferroglia, M.~Passera and G.~Passarino,
Nucl. Phys. B {\bf 777} (2007) pp.1-34, \par 35-99, 
100-156 (3 papers), 
DOI: 10.1016/j.nuclphysb.2007.(04.021, 03.043, 04.027).

\noindent
17) Stephen~P.~Martin and  David~G.~Robertson,
Phys. Rev. D {\bf 90}, 073010 (2014), \par
DOI: 10.48550/arXiv.1407.4336(revised version).

\noindent
18) M.~D.~Goodsell and S.~Pa\ss ehr, 
Eur. Phys. J. C (2020), 80:417, \par
DOI: 10.1140/epjc/s10052-020-7657-8.

\noindent
19) N.Nakanishi, Progr.Theor.Phys., {\bf 17}, (1957), pp.401-418.

\noindent
20) P.Cvitanovi\'{c} and T.Kinoshita, Phys.Rev. D {\bf 10}, (1974), pp.3978-4031(3 papers).

%==============
\end{document}